\documentclass[fleqn,10pt]{wlscirep}
\usepackage[utf8]{inputenc}
\usepackage{bm}

\usepackage{epsfig}
\usepackage{graphicx}
\usepackage{psfrag}
\usepackage{amsmath,amssymb,psfrag,slashed,graphicx}
\usepackage{colordvi}
\usepackage{color}

\usepackage{slashed}
\usepackage{color}
\usepackage[normalem]{ulem}

\title{Proton spin after 30 years: what we know and what we don't?}

\author[1,*]{Xiangdong Ji}
\author[2,$\dagger$]{Feng Yuan}
\author[3,$\ddagger$]{Yong Zhao}
\affil[1]{Center for Nuclear Femtography, SURA, 1201 New York Ave. NW, Washington, DC 20005, USA;
Department of Physics, University of Maryland, College Park, MD 20742, USA}
\affil[2]{Nuclear Science Division, Lawrence Berkeley National Laboratory, Berkeley, CA 94720, USA}
\affil[3]{Physics Department, Brookhaven National Laboratory Bldg. 510A, Upton, NY 11973, USA}

\affil[*]{e-mail: xji@umd.edu}
\affil[$\dagger$]{e-mail: fyuan@lbl.gov}
\affil[$\ddagger$]{e-mail: yzhao@bnl.gov}

\begin{abstract}
More than three decades has passed since the European Muon Collaboration
published the first surprising result on the spin structure of
the proton. Much theoretical and experimental progress has been
made in understanding the origins of the proton spin. In this review,
we will discuss what we have learned so far, what are still missing,
and what we shall expect to learn from the upcoming
experiments including JLab 12 GeV and Electron-Ion Collider.
In particular, we focus on first principles
calculations and experimental measurements of the total
gluon helicity $\Delta G$, and quark and gluon orbital angular momenta.
\end{abstract}

\begin{document}

\flushbottom
\maketitle

\thispagestyle{empty}

\noindent \textbf{Key points:}
\begin{itemize}
\item{There are two established approaches to look at the compositions of the proton spin:
frame-independent spin structure, $\Delta\Sigma/2+L_q+J_g=\hbar /2$ (``Ji sum rule'') and
infinite-momentum-frame or parton spin structure, $\Delta\Sigma/2+\Delta G+\ell_q+\ell_g=\hbar/2$ (``Jaffe-Manohar sum rule'').}

\item{In the frame-independent approach, quark orbital $L_q$ and gluon angular momentum
contributions $J_g$ can be extracted from moments of generalized parton distributions.
Results from Jlab 6 GeV and HERMES experiments suggest a substantial
quark orbital contribution $L_q$.}

\item{In terms of partons, the quark and gluon helicity contributions, $\Delta\Sigma/2$ and $\Delta G$,
have a simple physical interpretation, and the result from RHIC-spin experiments
has provided first important constraint on the total gluon helicity $\Delta G$.}

\item{Development of large-momentum effective theory along with lattice QCD
simulations provides first-principles calculations of the spin
structure. The recent results on $\Delta \Sigma$, $L_q$, $J_g$, and $\Delta G$ have provided intriguing theoretical
pictures.}

\item{JLab 12 GeV program will provide much improved information on quark orbital
angular momentum $L_q$ and $\ell_q$. Future Electron-Ion Collider will provide high-precision measurements
on the gluon helicity $\Delta G$ and gluon angular momentum $J_g$ and $\ell_g$.}
\end{itemize}

\vspace{1cm}

\section{Introduction}

The proton is a spin-1/2 particle, thought to be fundamental
when discovered as a basic constituent of atomic nuclei by Rutherford
in 1917~\cite{Rutherford:1919fnt}. However, the subsequent measurement of its magnetic
moment~\cite{Stern} showed a significant deviation from the
Dirac value for a point-like object~\cite{Dirac:1928hu}. Ever since, the
proton substructure along with the origin of its spin and magnetic moment has
intrigued nuclear and particle physicists for nearly a century.

Every model of the proton ought to give an explanation
for its spin: from the Skyrme model~\cite{Skyrme:1962vh}, to
Gell-Mann and Zweig's quark model~\cite{GellMann:1964nj,Zweig:1981pd},
and to many other models proposed in the 70's and 80's~\cite{Bhaduri:1988gc,Thomas:2001kw}.
The simplest and most successful one has been the quark model
which inspired, among others, the discovery of quantum
chromodynamics~(QCD)~\cite{Fritzsch:1973pi},
the fundamental theory of strong interactions.
The non-relativistic quark model has an exceedingly simple
explanation for the proton/neutron spin and the
associate magnetic moments~\cite{Greenberg:1964pe}, as well as
their excitations~\cite{Isgur:1978wd}:
Three constituent quarks are all in the $s$-wave orbit,
and their spins couple to $1/2$ in a way consistent with the
SU(2)$_{\rm spin}\times$SU(3)$_{\rm flavor}$ combined
spin-flavor symmetry~\cite{Close:1979bt}.

The quark model picture was put under a straight test
through polarized deep-inelastic scattering (DIS) on a
polarized proton target~\cite{Hughes:1983kf}.
In 1987, the European Muon Collaboration (EMC)
delivered a measurement for the fraction of the proton
spin carried by quarks~\cite{Ashman:1987hv,Ashman:1989ig},
\begin{align}
\Delta \Sigma(Q^2\!=\!10.7{\rm GeV}^2) = 0.060 \pm 0.047\pm 0.069\,,
\end{align}
which is consistent with zero!
The EMC data also showed a significant deviation from the Ellis-Jaffe sum rule
for the $g_1(x,Q^2)$ structure function based on the quark picture~\cite{Ellis:1973kp}.
This result literally shocked the
physics community, and created the so-called proton spin ``crisis'' or
proton spin problem. The discrepancy has since inspired a large number of experimental
and theoretical studies which have been reviewed in a number of papers
~\cite{Filippone:2001ux,Bass:2004xa,Aidala:2012mv,Leader:2013jra,Ji:2016djn,Deur:2018roz}.
The most important lesson we have learned is that the underlying
theory for the proton structure, QCD, has a much more sophisticated
way to build up the proton spin.

QCD is fundamental and beautiful on the one hand, and
is sophisticated and defies simple ways to understand
on the other. For example, it is no longer feasible, or
we have failed so far, to come up with an entire quark and gluon wave function for the
proton and to check the content of various components. Therefore, we will
consider instead the so-called {\it sum rules} or decompositions of
the spin into various physical parts. This has been the main approach
to understand the origins of the proton spin so far.

This article is not a comprehensive review of hadron spin
physics. In particular, it is not meant to be an update on the
recent reviews~\cite{Aidala:2012mv,Deur:2018roz} which have done an
excellent job. Rather, it focuses
sharply on the questions related to the origins of the proton spin.
We mainly discuss issues like: does it make sense to talk about different parts
of the proton spin? What will be an interesting and physically meaningful
decomposition for the spin? To what extent do we believe that we can measure each
part experimentally? How can one calculate them
in fundamental theory and put the results under experimental tests?
We hope that, after 30 years of the EMC result, this article can
help the physics community at large to understand what we know now, what we don't,
and what we shall expect in the future. In particular, what
the Electron Ion Collider (EIC) will help to answer the fundamental
questions about the origins of the proton spin~\cite{Accardi:2012qut,Boer:2011fh}.

\section{Spin Structure in Sum Rules}

Without knowing the wave function,
the angular momentum (AM) or spin structure of
a composite system can be studied through
various contributions to the total.
Thus, to explore origins of the proton spin,
we can start from QCD AM operator
in terms of individual sources,
\begin{equation}
\vec{J}_{\rm QCD} =\sum_\alpha \vec{J}_\alpha \ ,
\end{equation}
through which, the spin projection
$\hbar/2$ can be expressed as a sum of different
contributions.

One must be aware of some limitations in
this approach. Since the proton is an eigenstate
of the relativistic Pauli-Lubanski spin~\cite{Tung:1985na}, the
individual contributions can only be the quantum mechanical
expectation values of the AM sources from the
entire bound-state wave function. Moreover, they are ``renormalization-scale
dependent'', because individual operators are not separately conserved,
and the resulting ultra-violet (UV) divergences must be
renormalized in the senses that the short distance physics
is included in the effective AM operators~\cite{Ji:1995cu}. In non-relativistic systems,
with the exception of particles moving in a magnetic field,
the AM sources corresponding to different physical degrees of freedom
obey the separate AM commutation relations. In quantum field theories,
the simple commutation relations at the bare-field level are violated when
dressed with interactions, and only the total AM commutation relations
are protected by rotational symmetry~\cite{Ji:2010zza}. Finally, gauge symmetry
imposes important constraints on what is physically measurable.

Still there exist more than one way to split the AM operator and derive spin
sum rules for the proton. A physically-interesting spin sum rule shall have the
following properties:
\begin{itemize}
	\item{{\it Experimental Measurability}. The overwhelming interest in the proton spin
		began with the EMC data. Much of the followup experiments,
		including HERMES and COMPASS, polarized RHIC~\cite{Bunce:2000uv},
		Jefferson Lab 12 GeV upgrade~\cite{Dudek:2012vr} and EIC~\cite{Accardi:2012qut,Boer:2011fh}, have been
		partially motivated to search a full understanding of the proton spin.}
	\item{{\it Frame Independence:} Since spin is an intrinsic property of a particle,
		one naturally searches for a description of its structure independent of its momentum.
		How the individual contributions depend on the reference
		frame requires understanding on the Lorentz transformation properties of $\vec{J}_\alpha$.
Since the proton structure probed in high-energy scattering
is best described in the infinite momentum frame (IMF),
a partonic picture of the spin is interesting in this special frame of reference.}
\end{itemize}
According to these remarks, two sum rules have been well established in the literature (Table~\ref{Fig:sumrules}): the
frame-independent one~\cite{Ji:1996ek} and IMF one~\cite{Jaffe:1989jz}, as we explain below.

\subsection{QCD sources of angular momentum}

To obtain a spin sum rule, we need an expression
to the QCD AM operator. It can be
derived through Noether's theorem~\cite{Noether:1918zz} based on space-time
symmetry of QCD lagrangian density,
\begin{equation}
{\cal L}_{\rm QCD} = -\frac{1}{4} F^{\mu\nu}_a F_{\mu\nu a} + \sum_f \overline{\psi}_f (i\slashed D -m_f)\psi_f \ ,
\end{equation}
where $F^{\mu\nu}_a$ is a gluon field strength tensor or simply gluon field with color
indices $a=1,...,8$ and $\psi_f$ a quark spinor field of flavor $f=u,d,s,...$. The relation
between the gauge field and gauge potential $A^\mu_a$ is, $F^{\mu\nu}_a = \partial^\mu A^\nu
-\partial^\nu A^\mu_a -gf^{abc}A^\mu_b A^\nu_c$, and the covariant derivative
is $D^\mu = \partial^\mu +ig A^\mu$, with $A^\mu = A^\mu_a t^a$, and $t^a$
are the generators of the SU(3) color group and $f^{abc}$ are the structure constant.
Straightforward calculation yields the canonical AM expression~\cite{Jaffe:1989jz}
\begin{align}
\vec{J}_{\rm QCD} =& \int d^3 \vec{x}\left[\psi^\dagger_f\frac{\vec{\Sigma}}{2} \psi_f
+ \psi^\dagger_f \vec{x} \times (-i\vec{\partial})\psi_f  + \vec{E}_a\times \vec{A}_a + E^i_a(\vec{x}\times \vec{\partial})A^i_a\right] \ ,
\label{eq:amcanonical}
\end{align}
where $\vec{\Sigma}={\rm diag}(\vec{\sigma},\vec{\sigma})$ with $\vec{\sigma}$ being the Pauli matrix, and the contraction of flavor ($f$) and color ($a$) indices, as well as the spatial Lorentz index ``$i$'', is implied. The above expression
contains four different terms, each of which has clear physical
meaning in free-field theory. The first term corresponds to the quark spin,
the second to the quark orbital AM (OAM), the third to the gluon spin,
and the last one to the gluon OAM. Apart from the
first term, the rest are not gauge-invariant under the general
gauge transformation, $\psi\rightarrow U(x)\psi$ and $A^\mu \rightarrow U(x)\left(A^\mu
+ (i/g)\partial^\mu\right)U^\dagger(x)$, where $U(x)$ is an SU(3) matrix.
However, the total
is invariant under the gauge transformation up to a surface
term at infinity which can be ignored in physical matrix elements.

Theoretically, the canonical form of the AM operator allows deriving
an infinite number of spin sum rules with choices of gauges
and/or frames of reference (hadron momentum)~\cite{Ji:2012gc,Leader:2013jra}.
In practice, only the infinite-momentum frame, relevant for interpreting high-energy scattering
experiments, and physical gauge, such as Coulomb gauge, have shown related
to experimental observables.

Using the Belinfante improvement procedure~\cite{Belinfante:1940},
one can obtain a gauge-invariant form from Eq. (\ref{eq:amcanonical})~\cite{Ji:1996ek},
\begin{align}
\vec{J}_{\rm QCD}=&  \int d^3 x \left[ \psi^\dagger_f \frac{\vec{\Sigma} }{2}\psi_f +
\psi^\dagger_f \vec{x} \times (-i\vec{\partial}-g\vec{A}) \psi_f  +  \vec{x}\times(\vec{E}\times\vec{B})\right]\,,
\label{eq:amgi}
\end{align}
All terms are manifestly gauge independent, with the second term as mechanical or kinetic
OAM, and the third term gluon AM.

To evaluate the quark orbital and gluon contributions in a polarized proton state, we need the matrix elements
of the QCD energy-momentum tensor (EMT), which can be slit into the sum of the quark and gluon
contributions, $T^{\mu\nu}=T^{\mu\nu}_q+T^{\mu\nu}_g $, after Belinfante improvement.
EMT defines the momentum density which is the source of AM density. The off-forward matrix elements of
EMT have been parameterized as~\cite{Ji:1996ek},
\begin{align}
 \langle P'S|T^{\mu\nu}_{q/g}(0)|PS\rangle &=\bar U(P'S) \Big[A_{q/g}(\Delta^2)\gamma^{(\mu}\bar P^{\nu)}  +B_{q/g}(\Delta^2)\frac{\bar P^{(\mu}i\sigma^{\nu)\alpha}\Delta_\alpha}{2M}
+ ...\Big]U(PS) \  , \label{energy}
\end{align}
where $\bar P^\mu=(P^\mu+{P'}^\mu)/2$, $\Delta^\mu={P'}^\mu-P^\mu$. $U$ and $\bar{U}$ are Dirac spinors for the nucleon state, and $A$, $B$ are form factors depending the momentum transfer squared, $\Delta^2$.

\subsection{Helicity sum rules}

Without loss of generality, one can assume the proton three-momentum to
be $\vec{P} = (0,0,P^z)$. In the case of longitudinal polarization, one
has $ \langle PS_z|J^z|PS_z\rangle = {\hbar}/{2}$
where $S_z$ is spin polarization vector. The above equation
is boost-invariant along the $z$-direction. This
is a starting point to construct helicity (projection of the spin along the direction
of motion) sum rules.

Using the gauge-invariant
QCD AM in Eq.(~\ref{eq:amgi}),
one can can write down the frame-independent sum rule~\cite{Ji:1996ek,Ji:1997pf},
\begin{equation}
\frac{1}{2}\Delta \Sigma(\mu)  + L^z_q(\mu) + J_g(\mu)=\frac{\hbar}{2}\,, \label{eq:ji}
\end{equation}
where $\Delta\Sigma/2$ is the
quark helicity contribution measured in the EMC experiment, and $L^z_q$ is
total quark OAM contribution including all flavors of quarks.
Together, they give the total quark AM contribution $J_q$. The last term, $J_g$,
is the gluon contribution. All contributions depend on renormalization scheme
and scale $\mu$, which are usually taken to be dimensional regularization and (modified)
minimal subtraction. It has been shown that both contributions are related to
the form factors of the energy momentum tensor, $J_{q,g}=[A_{q,g}(0)+B_{q,g}(0)]/2$~\cite{Ji:1996ek}.

The frame-independence of the above sum rule means that the proton spin composition
does not depend on its momentum so long its helicity is a good quantum number, be it in
the finite momentum frame or infinite momentum frame.
This is a nice feature because
the wave function is clearly frame dependent.

Helicity sum rules can also be derived from the canonical expression
of the QCD AM density in Eq.~(\ref{eq:amcanonical}).
Because of gauge dependence, one might outright dismiss
the physical relevance of such sum rules.
However, as we shall explain in the next subsection that the gluon helicity
contribution in the IMF is actually physical. This prompts speculations
that the quark and gluon canonical OAM
might be measurable as well under such condition. Therefore,
Jaffe and Manohar proposed a canonical spin sum rule in a nucleon
state with $P^z=\infty$~\cite{Jaffe:1989jz},
\begin{equation}
\frac{1}{2} \Delta \Sigma(\mu) + \Delta G(\mu) + \ell_q(\mu) + \ell_g(\mu) = \frac{\hbar}{2} \ , \label{eq:jaffe-manohar}
\end{equation}
where $\Delta G$ is the gluon helicity and $\ell_{q,g}$ are
the canonical quark and gluon OAM, respectively.
Considerable studies have been made about this sum rule in the literature
because of its relevance to parton physics of the proton.
A recent precision study of renormalization scale $\mu$ dependence of $\Delta \Sigma$ and $\Delta G$ has been
reported in Ref.~\cite{deFlorian:2019egz}, see also \cite{Ji:1995cu} for the scale evolution of OAM
contributions.

Recent developments~\cite{Hatta:2011ku,Lorce:2011kd,Lorce:2011ni,Ji:2012sj} have shown that the parton OAM can be closely connected to the quantum phase space Wigner function or distribution~\cite{Ji:2003ak,Belitsky:2003nz}.
Since the Wigner function describes the quantum distribution of quarks and gluons in both spatial and momentum spaces, we can construct the parton OAM by a properly-weighted integral. This leads to an intuitive explanation of
the OAM contributions in the above two sum rules. The difference between two
OAM's, the so-called potential angular momentum, comes from two different ways to define gauge links in the Wigner functions~\cite{Ji:2012sj}, one of which can be interpreted as final-state interaction effects in scattering experiments~\cite{Wakamatsu:2012ve,Burkardt:2012sd}.

There has been other attempts in using Eq.~(\ref{eq:amcanonical})
to write down sum rules in different frames and gauges, for example,
the Coulomb gauge at finite hadron momentum~\cite{Chen:2008ag}. However, these sum rules have no known
experimental measurements and remain a pure theoretical
interest~\cite{Ji:2012gc,Leader:2013jra}. Some of them are known to
reduce to the Jaffe-Manohar sum rule in the IMF~\cite{Ji:2013fga,Hatta:2013gta}.

\begin{table}
	\begin{center}
		\includegraphics[width=0.7\textwidth]{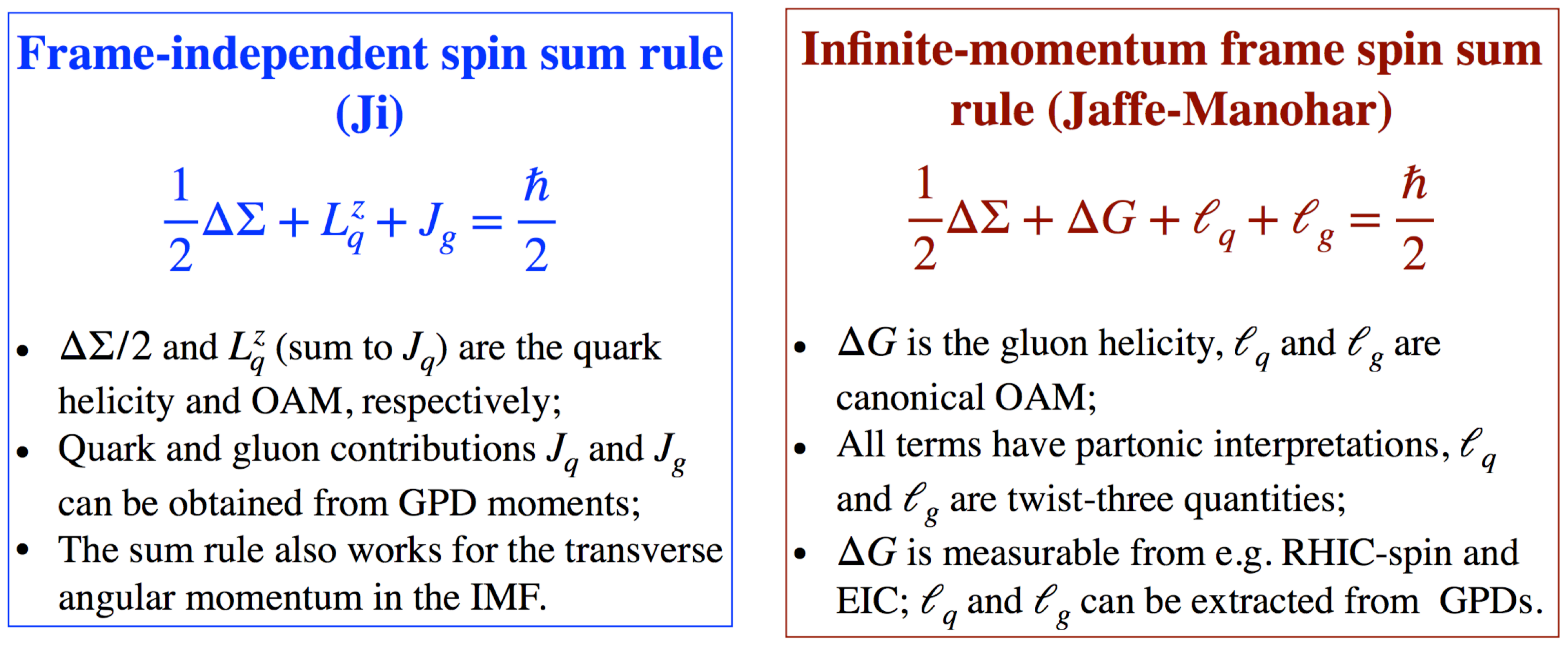}
		\caption{Two established proton spin sum rules: one is frame-independent~\cite{Ji:1996ek} and the other is
in the infinite-momentum frame~\cite{Jaffe:1989jz}.}
		\label{Fig:sumrules}
	\end{center}
\end{table}

\subsection{Why is gluon helicity in bound states a physical quantity?}

In general, a gauge-dependent operator is not a physical observable
and hence cannot be related to an experimental measurement.
However, $\Delta G$ and OAM in IMF in Eq. (\ref{eq:jaffe-manohar})
appear exceptional. This has been an interesting theoretical puzzle for many years,
and has generated much debated in the literature~\cite{Hoodbhoy:1998bt,Hoodbhoy:1999dr,Chen:2008ag,Ji:2012gc,Hatta:2011zs}.

Experimentally-measurable $\Delta G$ is the first moment of the gauge-invariant
polarized gluon distribution~\cite{Manohar:1990jx},
\begin{align}
\label{eq:deltag}
\Delta G(Q^2) = \int_0^1 dx\ \Delta g(x,Q^2)\,,\ \ \ \ \ \ \Delta g(x)=  {i\over 2x(P^+)^2} \int {d\lambda\over 2\pi} e^{i\lambda x} \langle PS| F^{+\alpha} (0) W(0,\lambda n)\tilde{F}^{~+}_{\alpha}(\lambda n)|PS\rangle  \,,
\end{align}
where $\tilde{F}^{\alpha\beta} = \epsilon^{\alpha\beta\mu\nu}F_{\mu\nu}/2$, and the light-cone gauge link $W(0,\lambda n)$ is defined in the adjoint representation of SU(3). [IMF quantities have been rewritten in
the standard light-front notation $V^\pm = (V^0\pm V^z)/\sqrt{2}$ after Lorentz transformation.] The above
quantity clearly is gauge-invariant, but non-local. It does not seem to have a simple
interpretation in a general gauge.

However, in the light-cone gauge $A^+=0$, the nonlocal operator in Eq.~(\ref{eq:deltag})
reduces to the gluon ``spin operator'' in Eq. (\ref{eq:amcanonical}), and thus the
experimental data on $\Delta G$ can be interpreted as the measurement of
a contribution to the Jaffe-Manohar's spin sum rule in this particular gauge.
This suggests that a gauge-variant operator
might correspond to an experimental observable in a specific gauge!
The latter has inspired much discussions about the gauge symmetry and
myriads of experimentally-unaccessible spin sum rules~\cite{Leader:2013jra}.
The fundamental reason is, however, not about generalizing the
concept of gauge invariance, but about the nature of the proton states
in the IMF~\cite{Ji:2013fga}.

As realized by Weizs\"{a}cker and
Williams in electromagnetism~\cite{vonWeizsacker:1934nji,Williams:1935dka},
the gauge field strength in a fast moving source is dominated
by its transverse components. For a static charge, the electric field
is purely longitudinal ($\vec{E}=\vec{E}_\parallel$ or without curl).
As the charge moves with velocity $\beta=v/c$, where $c$ is the speed of light,
the field lines start to contract in the transverse
direction due to Lorentz transformation.
The moving charge forms an electric current that generates transverse magnetic fields,
\begin{equation}
\vec{B}=\vec{\nabla}\times\vec{A} = \vec{\nabla}\times\vec{A}_\perp\,,
\end{equation}
and the gauge potential $\vec{A}$ acquires a non-zero transverse component $\vec{A}_\perp$ (divergence free).
At large $\beta$, the field strength gets enhanced by a factor of $\beta\gamma$ $(\gamma=1/\sqrt{1-\beta^2})$ in the transverse direction,
whereas it is strongly suppressed in the longitudinal direction~\cite{vonWeizsacker:1934nji,Williams:1935dka}. In the limit of $\beta\to1$ (or $\gamma\to\infty$), $\vec{E}_\perp\sim\vec{B}$, and $|\vec{E}_\perp|\gg|\vec{E}_\parallel|$, so the electromagnetic field can be approximated as free radiation!

The radiation fields have only two physical degrees of freedoms, and
the longitudinal one in the gauge potential is just a pure gauge. Thus,
for an on-shell photon, its helicity is physical and can be considered as
gauge-invariant spin. One can superimpose such on-shell plane wave
states with definite helicity to construct light modes
with definite OAM or so-called twisted light~\cite{Allen:1992zz,Bliokh:2015doa}.
The gauge-invariant issue never arises because one deals with physical polarization
at all time.

Analogously, the Weizs\"{a}cker-Williams approximation is also a valid picture
for gluons in an ultra-relativistic proton~\cite{Altarelli:1977zs}. In the IMF,
the gluons can also be approximated as free radiation, thus it only has two
physical transverse polarizations.
$A^+=0$ is a physical gauge which leaves the transverse
polarizations of the radiation field intact. This justifies
$\vec{E}\times\vec{A}=\vec{E}_\perp\times\vec{A}_\perp$ as the
physical gluon spin (helicity)
operator in the Jaffe-Manohar sum rule. The above
consideration also applies to the associated
canonical OAM, $\ell^z_q$ and $\ell^z_g$, which implies
partonic sum rules from them~\cite{Hagler:1998kg,Harindranath:1998ve,Bashinsky:1998if}.
However, the associated canonical OAMs involves transverse momentum integral by construction and their scale evolutions are much more complicated~\cite{Hoodbhoy:1998yb,Hatta:2019csj}.

The situation is quite different, however, if one considers
color fields inside a bound state which does not travel relativistically.
The longitudinal gauge potential subjected to gauge transformation
now contains a physical component whose effects cannot be separated
from the transverse part. The gluons are now off-mass-shell,
and the longitudinal polarization do has physical significance.
Only gauge-invariant operators can pick up the correct physics
from the longitudinal part of gluon potential.

Thus it is the physical states in IMF which ensure the total gluon
helicity is measured through $\vec{E}\times \vec{A}$.
The spin operator can have any longitudinal pure gauge potential which
does not contribute to the physical matrix element.
This situation is exactly opposite to the usual textbook formulation of gauge
symmetry where the external states are gauge-dependent and the operators
must be gauge invariant. When transforming the IMF states
into ones with finite momentum through infinite Lorentz transformation,
$\vec{E}_\perp\times\vec{A}_\perp$ becomes a non-local one in Eq. (\ref{eq:deltag}).

\subsection{Sum rule for transverse angular momentum}

For transverse polarization along, e.g., the $x$-direction,
the transverse AM operator does not commute with the QCD
Hamitonian. However, according to the Lorentz transformation
property of $J^x$, its expectation value in transversely-polarized
state is well-defined~\cite{Ji:2020hii},
\begin{equation}
\langle PS_x |J^x |PS_x \rangle  = \gamma ({\hbar}/{2})  \ .
\end{equation}
where $\gamma$ is the Lorentz boost factor. Therefore,
the transverse AM $J^x$ is a leading observable
because it enhances under boost, a fact less
appreciated in the literature. The potential
contribution to the transverse AM from the non-intrinsic
center-of-mass motion has led to
incorrect results in the literature~\cite{Bakker:2004ib,Leader:2011cr,Ji:2020hii}.

If we define, $
J^{q,g}_\perp =
\langle PS_\perp|J_{\perp q,g}|PS_\perp\rangle /(\gamma s_\perp)$,
then the quark and gluon contributions can again
be related to the form factors in Eq.~(\ref{energy}),
\begin{align}
& J^{q, g}_\perp = (A_{q,g}+B_{q,g})/2 ,  \\
& J^q_\perp + J^g_\perp = \hbar/2 \ .
\label{eq:transversesumrule}
\end{align}
Both equations are the same as these in the helicity case. However,
the separation of the quark spin and orbital contributions are frame dependent, with
the former contribution going to zero in the infinite momentum limit~\cite{Ji:2020hii}.

A parton interpretation can be derived for the above result following an earlier suggestion in Ref.~\cite{Burkardt:2002hr,Burkardt:2005hp}. The physical reason
is that the transverse AM can be built from a longitudinal parton
momentum with a transverse coordinate.
One can define a parton AM
partonic density~\cite{Hoodbhoy:1998yb,Ji:2012sj,Ji:2012vj},
\begin{equation}
J^{q}_\perp (x) = x\left[q(x)+ E_{q}(x)\right]/2, ~~~J^{g}_\perp (x) = x\left[g(x)+ E_{g}(x)\right]/2,
\end{equation}
where $q(x)$ and $g(x)$ are the unpolarized quark/antiquark and gluon distributions,
and $E_{q,g}(x)$ are a type of generalized parton distributions (GPDs)~\cite{Ji:1996ek}.
GPD's are an extension of the well-known Feynman parton distribution and are defined as
off-forward matrix element between nucleon states with different momenta, similar
to form factors. They depend on three kinematic variables: $x$ the longitudinal momentum fraction for the parton, $\xi$ the skewness parameter represents the momentum transfer between the nucleon states along the longitudinal direction, and $t=\Delta^2$ the momentum transfer $\Delta^\mu$ squared. They can be systematically studied through a new class of exclusive hadronic reactions~\cite{Ji:1996ek}.
$J^{q,g}_\perp(x)$ are the AM densities carried by partons of momentum
$x$ in a transversely polarized nucleon in which partons are in general off the center of mass
~\cite{Burkardt:2005hp}. Integrating the above over $x$
give the total transverse AM carried by quarks and gluons, respectively.

\section{{\it \bf Ab Initio} Calculations of Spin in Lattice QCD}

Because non-perturbative QCD is unusually challenging, a large number of
models of the proton have been proposed in 70's and 80's,
many of which use ``effective''
degrees of freedom. An introduction about
these models can be found in the textbooks~\cite{Bhaduri:1988gc,Thomas:2001kw}.
A recent one is the holographic model in which the proton
is pictured as a quark and di-quark bound state~\cite{Brodsky:2014yha}.
Since the connections between the model
degrees of freedom and the fundamental ones are unknown, whereas high-energy
experiments probe QCD quarks and gluons directly, we will
discuss the theoretical calculations using QCD degrees of freedom only.

At present, the only systematic approach
to solve the QCD proton structure is lattice field theory~\cite{Wilson:1974sk},
in which quark and gluon fields are put on four-dimensional Euclidean lattices with
finite spacing $a$, and quantum correlation functions of fields
are calculated using Feynman path integrals and Monte Carlo simulations.
The physical limits are recovered
when the lattice spacing $a$ becomes sufficiently small compared to physical
correlation length, the lattice volume much larger than hadron sizes,
and the quark masses close to the physical ones~\cite{Rothe:1992nt}.
There are less systematic approaches
such as Schwinger-Dyson (Bethe-Salpeter) equations~\cite{Maris:2003vk}
and instanton liquid models~\cite{Schafer:1996wv} in which a certain
truncation is needed to find a solution.
Although much progress has been made
in these other directions, we focus on the lattice QCD method which
can be systematically improved.

A complete physical calculation on lattice faces a number of obstacles.
First, the total AM is a flavor-singlet quantity, and as such,
one needs to compute the costly disconnected diagrams for the quarks.
Since up and down quarks are light, computation demands at the physical
pion mass are very high, as physical propagators becomes singular in the massless limit.
Moreover, gluon observables need be calculated to complete the picture,
which are known to be very noisy and large number of field configurations
are needed for accuracy. At the same time, one needs to take the continuum
and infinite volume limits. All of these add up to an extremely
challenging task. However, a computation with all these issues taken into account
has become feasible recently~\cite{Alexandrou:2020sml}.

An additional challenge
is present in quantities like $\Delta G$, usually defined in terms of
light-front correlations with real time dynamics. It is well-known that
the real-time Monte carlo simulations demand exponentially-increasing
resources. The recent development
in large-momentum effective theory (LaMET) has opened the door for such
time-dependent light-front correlations~\cite{Ji:2013dva,Ji:2014gla,Ji:2020ect}.

\subsection{Frame-independent helicity sum rule}

The matrix elements of local operators, $\Delta \Sigma$,
$J_q$ and $J_g$, are relatively easier to calculate using the
standard lattice QCD technique. Much progress has been made in
understanding the content of manifestly gauge-invariant
helicity sum rule (and hence the transverse AM sum rule as well
by Eq.~(\ref{eq:transversesumrule})).

The first calculation has been about the $\Delta \Sigma$
from different quark flavors~\cite{Dong:1995rx}. The relevant studies in the last two decades
have been summarized in a recent review~\cite{Lin:2017snn}.
Important progress has been made in chiral-fermion calculations~\cite{Liang:2018pis}
and at the physical quark mass~\cite{Alexandrou:2017oeh}.
The strange quark contribution has been calculated earlier
in~\cite{Deka:2013zha,Gong:2015iir} in consideration of the anomalous Ward identity.
The total quark spin contribution to the proton helicity has been found consistently
about 40\%.

The calculation of the total quark and gluon angular momenta
started in Ref.~\cite{Mathur:1999uf} where the quark part
including the disconnected diagrams was calculated in the quenched approximation.
The result of the total quark contribution is $J_q = 0.30\pm 0.07$, i.e. 60\%.
Therefore about 40\% of the proton spin is carried by gluon through
simple sum rule deduction. Following the quenched studies~\cite{Hagler:2003jd,Gockeler:2003jfa},
dynamical simulations have now become a standard~\cite{Brommel:2007sb,Bratt:2010jn,Syritsyn:2011vk,Alexandrou:2011nr,Alexandrou:2013joa}.
A first complete study of the AM decomposition was made in Ref.~\cite{Deka:2013zha},
followed by a chiral dynamical simulation recently~\cite{Yang:2019dha}. A first study
at the physical quark mass has appeared in Ref.~\cite{Alexandrou:2017oeh}.

\begin{figure}[tbp]
	\begin{center}
		\includegraphics[width=0.34\textwidth]{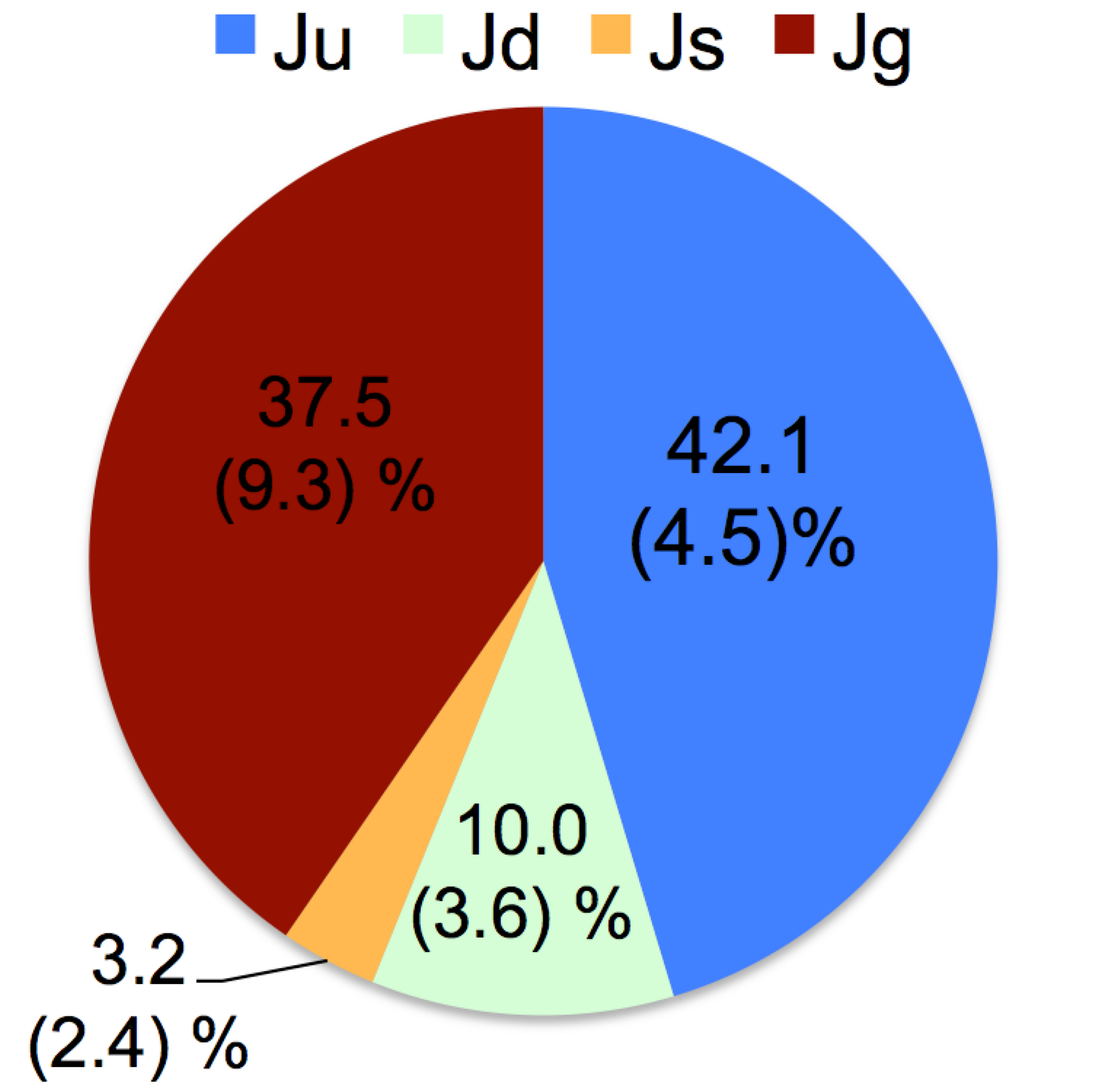}
		\includegraphics[width=0.5\textwidth]{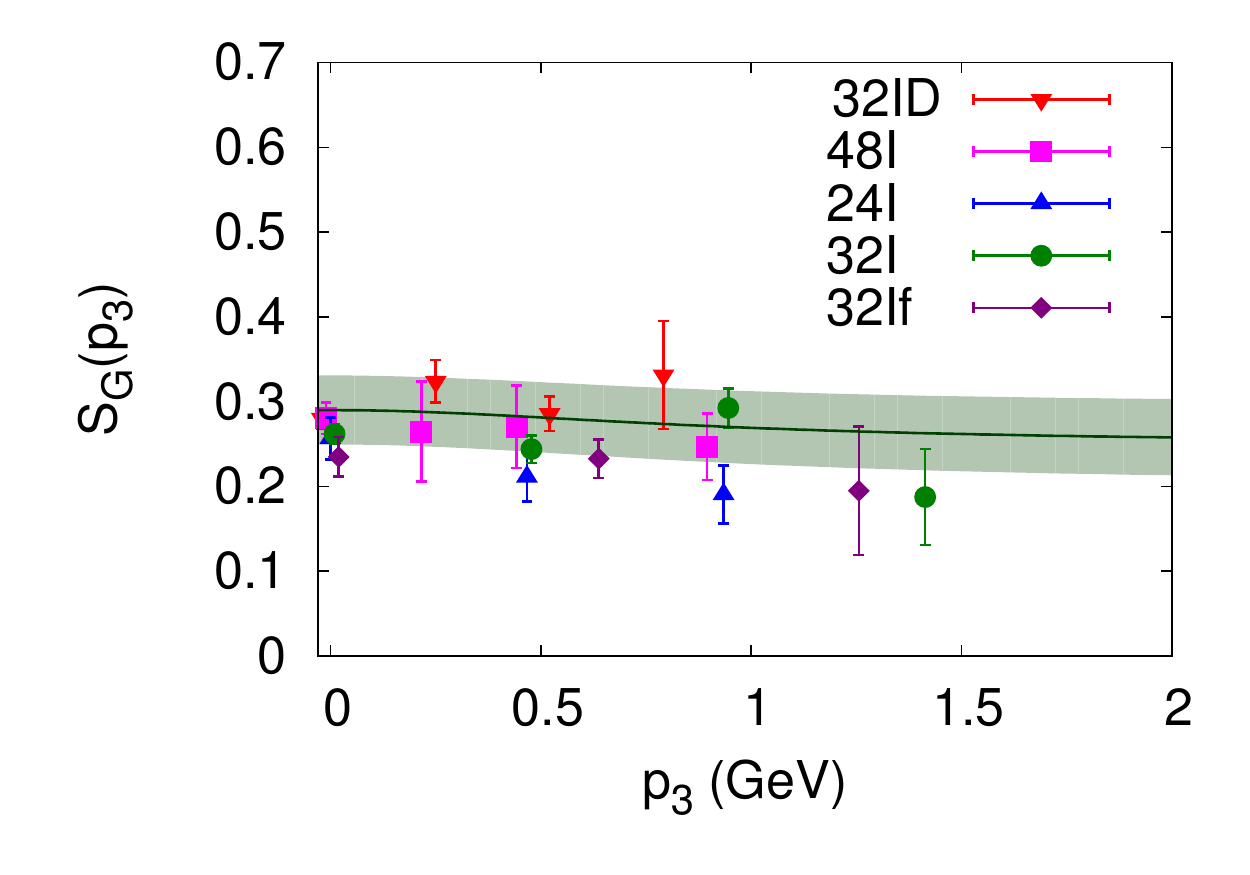}
		\caption{State-of-the-art lattice study on the proton spin. Left chart shows the spin decomposition in the frame-independent sum rule: total spin contributions from the up ($J_u$), down ($J_d$), strange ($J_s$) quarks and gluons ($J_g$). The numbers come from the ETMC collaboration~\cite{Alexandrou:2020sml} with $=38.2(3.1)\%$ from total quark helicity contribution $\frac{1}{2}\Sigma$ and $=18.8(10.1)\%$ from total quark orbital angular $L_q$.
Right plot shows the gluon helicity contribution to the IMF sum rule, where the gluon spin is
computed from different proton momenta (labeled as $p_3$ as $x$-axis) and lattice ensembles (different volumes and lattice spacing noted in the plot legend) by the $\chi$QCD collaboration~\cite{Yang:2016plb}.
The gluon spin reduces to $\Delta G$ when extrapolated to the IMF ($p_3\to \infty$): $\Delta G=0.251(50)$ or $50(9)\%$ of the proton spin, which can be compared to the RHIC-spin determination, see, Fig.~\ref{fig:gluonspin}. }
		\label{fig:emtc}
	\end{center}
\end{figure}

A high-precision dynamical simulation at the physical pion mass has
been finished recently~\cite{Alexandrou:2020sml}. It was found that
the total quark spin contribution is about $38.2\%$, and the orbital AM
of the quarks about 18.8\%, much reduced compared with
quenched simulations. The total gluon contribution
is 37.5\%. The resulting pie chart is shown in Fig.~\ref{fig:emtc}. The total
spin is 94.6\% of $\hbar/2$ with an error bar of 14.2\%. These results are
largely consistent with the chiral fermion study in Ref.~\cite{Yang:2019dha}.
All numbers are quoted in $\overline{\rm MS}$ scheme at $\mu=2 $ GeV.

\subsection{Gluon helicity in light-travelling proton}

Calculation of the gluon helicity $\Delta G$ has not been possible for
many years because it is intrinsically a time-dependent light-front quantity.
However, a breakthrough in 2013 has finally been made by
studying the frame dependence of non-local matrix elements~\cite{Ji:2013fga}.
It was found that one can match the
large-momentum matrix element of a static ``gluon spin'' operator calculable in lattice QCD
to $\Delta G$ in the IMF~\cite{Ji:2013fga}. This idea was a prototype of LaMET,
which was soon put forward as a general approach to calculate all parton physics~\cite{Ji:2013dva,Ji:2014gla}.

The choice of the static ``gluon spin'' operator is not unique. There is a universality class of operators~\cite{Hatta:2013gta} whose IMF limit approach the free-field field operator
in Eq.~(\ref{eq:amcanonical}) in the light-front gauge.
The simplest choice for the static ``gluon spin'' is the free-field operator $\vec{E}\times \vec{A}$ fixed in a time-independent gauge. For example, the Coulomb gauge $\vec{\nabla}\cdot \vec{A}=0$, axial gauges $A^z=0$ and $A^0=0$ all maintain the transverse polarizations of the gluon field in the IMF limit, so they are viable options.

In the Coulomb gauge and $\overline{\rm MS}$ scheme, the static ``gluon spin'' $\Delta \widetilde G$ in a massive on-shell quark state at one-loop order is~\cite{Chen:2011gn,Ji:2013fga}
\begin{align}  \label{eq:coulomb}
\Delta \widetilde G(P^z, \mu)  (2S^z) &=\left.\langle PS| (\vec{E}\times \vec{A})^z|PS\rangle_q \right\arrowvert_{\vec{\nabla}\cdot \vec{A}=0} = \frac{\alpha_sC_F}{4\pi} \left[ {5\over 3}\ln{\mu^2\over m^2} -\frac{1}{9} + \frac{4}{3}\ln \frac{(2P^z)^2}{m^2}\right](2S^z) \,,
\end{align}
where the subscript $q$ denotes a quark. The collinear divergence is regulated by the finite quark mass $m$.
The above result shows that the gluon state depends on the three-momentum $P^z$, as it should be.
If we follow the procedure in~\cite{Weinberg:1966jm} and take $P^z\to\infty$ limit before UV regularization, which is the standard procedure to define partons~\cite{Ji:2013fga},
\begin{align} \label{eq:coulombimf}
\Delta G(\infty, \mu) (2S^z)&= \left. \langle PS|(\vec{E}\times \vec{A})^z|PS\rangle_q\right\arrowvert_{\vec{\nabla}\cdot \vec{A}=0} =\frac{\alpha_sC_F}{4\pi}\left(3\ln{\mu^2\over m^2}+7\right) (2S^z)\,,
\end{align}
which is exactly the same as the light-front gluon helicity $\Delta G(\mu)$
appeared in Jaffe-Manohar spin sum rule~\cite{Hoodbhoy:1998bt}. Therefore,
despite the difference in the UV divergence, the infrared-sensitive collinear divergences
of $\Delta \widetilde G(P^z,\mu)$ and $\Delta G(\mu)$ are exactly the same,
which allows for a perturbative matching between them.

With the LaMET approach, $\Delta G$ was calculated in lattice QCD for the first time~\cite{Yang:2016plb}. In this calculation, the static gluon spin operator $\vec{E}\times \vec{A}$ in the Coulomb gauge was simulated on the lattice and converted to the continuum $\overline{\rm MS}$ scheme with one-loop lattice perturbation theory, which is shown in Fig.~\ref{fig:emtc}. With leading-order matching and extrapolation to the IMF, the authors obtained $\Delta G(\mu=\sqrt{10}\mbox{ GeV})=0.251(47)(16)$, or $50(9)(3)\%$ of the proton spin.
A refined study on systematics and precise matching shall be made in the future.

\subsection{Canonical OAM and Transverse AM density in light-travelling proton}

To complete the Jaffe-Manohar picture of the proton spin, one needs to
compute canonical OAM of the quarks and gluons in the IMF and light-front gauge.
This can be done following the same approach above for $\Delta G$.
A study of calculating these in LaMET has been made in Ref.~\cite{Zhao:2015kca}. One
can start from the matrix elements, for example, in
Coulomb gauge and at finite momentum $P^z$,
\begin{align}
\tilde \ell_q(\mu,P^z)(2S^z) &= \langle PS| \int d^3\vec{x}\ \psi^\dagger_q (\vec{x}\times (-i\vec{\nabla}))^z\psi_q |PS\rangle \,,\ \ \ \ \ \ \tilde \ell_g(\mu,P^z)(2S^z) = \langle PS| \int d^3\vec{x}\  E^{ia} (\vec{x}\times \vec{\nabla})^zA^i_a |PS\rangle\,,
\end{align}
which can be matched onto $\ell_{q,g}(\mu)$ in the Jaffe-Manohar
sum rule. The matching expressions have been worked out
in Coulomb gauge in Ref.~\cite{Ji:2014lra}. Mixings with
potential AM contributions shall be taken into account~\cite{Wakamatsu:2012ve}.
Because the matrix elements are spatial moments, one can either
calculate them directly using $\vec{x}$-weighting on
lattice~\cite{Gadiyak:2001fe,Blum:2015gfa},
or by taking the zero-momentum-transfer limit of the momentum-density
form factors. Computing the canonical quark OAM from lattice QCD has been
carried out in Ref.~\cite{Engelhardt:2017miy,Engelhardt:2020qtg} using non-local operators,
for which matching to the IMF quantities has yet to be studied.

Similar approach can be used to calculate the canonical OAM distributions $\ell_q(x,\mu)$ and $\ell_g(x,\mu)$~\cite{Hagler:1998kg,Bashinsky:1998if}.
Since both distributions are sub-leading in high-energy experiments (the so-called twist-three), they may contain
a zero-mode contribution at $x=0$~\cite{Aslan:2018tff,Ji:2020baz}, which makes
the experimental measurement of $\ell_{q,g}(\mu)$ through sum rules challenging.
Mixings with other twist-three correlations with gluon fields must be
considered.

Likewise, the transverse AM of the proton
is a leading light-front observable, and has a
partonic interpretation in terms of
transverse AM density $J^{q,g}_\perp(x)
= x(\{q,g\}(x)+E_{q,g}(x))/2$. While the
singlet distributions $q(x),g(x)$ are well constrained and can be
calculated on lattice with the standard LaEMT
method~\cite{Ji:2020ect}, little is known
about GPD's $E_{q,g}(x)$. The moments of $E_{q,g}(x)$ can be calculated as
a generalization of the form factors of the energy-momentum tensor.
The $x$-distributions can also be obtained directly as the spatial moment of
the gauge-invariant momentum-density correlation functions.

\section{Experimental Progress and Electron-Ion Collider}

 We finally review the experimental progress in searching for the origins of the proton spin. Following the EMC,
 many experiments were launched to confirm the result. In the first subsection, we discuss efforts of nailing
 down quark helicity contribution $\Delta q$, particularly $\Delta s$, from SIDIS, and the gluon helicity $\Delta G$ from polarized proton-proton collisions at RHIC. In the second subsection,
 we review measuring the quark orbital AM contribution from a new class of
 experiments called deeply-virtual Compton scattering, first proposed and studied
 in Refs.~\cite{Ji:1996ek,Ji:1996nm}. Following this, we consider
 the prospects of studying the proton spin structure at the EIC.

\subsection{Nailing down the quark and gluon helicities}

\begin{figure}[tbp]
	\begin{center}
		\includegraphics[width=0.80\textwidth]{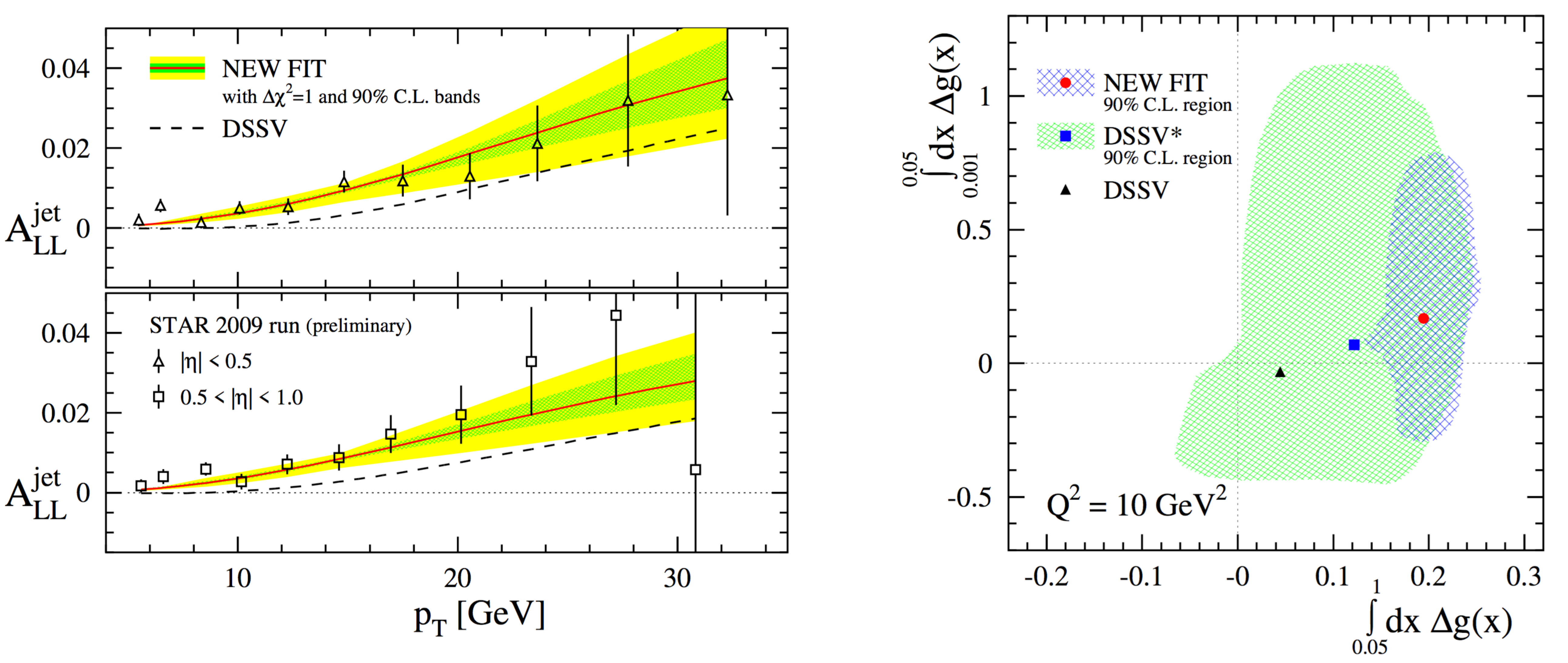}
			\end{center}
\caption{The relativistic heavy-ion collider (RHIC) at Brookhaven National Laboratory (BNL) provides a strong evidence of the gluon helicity contribution to the proton spin. (Left) Double spin asymmetry in inclusive jet production at RHIC compared to the the global analysis of DSSV-14~\cite{deFlorian:2014yva}, where the gluon helicity $\Delta g(x)$ plays an important role. (Right) Constraints on the gluon helicity contribution (labeled as ``NEW FIT") to the proton spin from the fit to the experimental data including that on the left plot. In the RHIC kinematics, i.e., $x>0.05$, $\Delta g$ was found positive and sizable: $\int_{0.05}^1 dx \Delta g(x)=0.20_{-0.07}^{+0.06}$, as shown in abscissa. However, in the unexplored region of $x<0.05$, the uncertainties are still significant as shown in ordinate. Source: experiment data from~\cite{Adamczyk:2014ozi}, DSSV-14 global analysis in Ref.~\cite{deFlorian:2014yva}.}
		\label{fig:gluonspin}
\end{figure}

The majority of the experiment efforts followed the EMC experiment, measuring the polarized structure functions in DIS  with polarized lepton on polarized target (proton, neutron, deuteron). Two important initiatives have also emerged: 1). The DIS experiment facilities extend their capabilities to measure the spin asymmetries in the semi-inclusive hadron production in
DIS (SIDIS)~\cite{Airapetian:2004zf,Alekseev:2009ac}, helping to identify the flavor dependence in the polarized quark distributions. 2). The Relativistic Heavy Ion Collider (RHIC) at the Brookhaven National Laboratory (BNL) started the polarized proton-proton scattering experiments, which opened new opportunities to explore the proton spin, in particular, for the helicity contributions from gluon and sea quarks. Most of these efforts have been covered in the recent reviews~\cite{Aidala:2012mv,Deur:2018roz}.

The total quark spin contribution has been well determined from DIS measurements: $\Delta \Sigma \approx 0.30$ with uncertainties around $0.10$, see, e.g., recent global analyses from Refs.~\cite{deFlorian:2009vb,deFlorian:2014yva,Nocera:2014gqa}. However, for sea quark polarizations including $\bar u$, $\bar d$ and $s$ ($\bar s$), there exist larger uncertainties, in particular, in the strange quark polarization~\cite{deFlorian:2014yva,Nocera:2014gqa,Ethier:2017zbq}, where the constraints mainly come from SIDIS measurements by HERMES and COMPASS experiments. Recently, it was also found that the $W$ boson spin asymmetries at center-of-mass energy $\sqrt{s}=510$ GeV at RHIC have also improved the constraints on $\bar u$ and $\bar d$ polarization~\cite{Adam:2018bam}. Very exciting results, in particular, for the double spin asymmetries in inclusive jet production from the RHIC experiments have provided stronger constraint on the gluon helicity~\cite{Adamczyk:2014ozi}, see Fig.~\ref{fig:gluonspin}. This promises great potential for future analysis from RHIC experiments to further reduce the uncertainties due to improved statistics~\cite{Aschenauer:2015eha,Adam:2019aml}. However, due to the kinematic limitations, the total gluon helicity contribution still has a significant uncertainty.

\subsection{Quark OAM and GPD studies at JLab 12 GeV}

\begin{figure}[tbp]
	\begin{center}
		\includegraphics[width=0.8\textwidth]{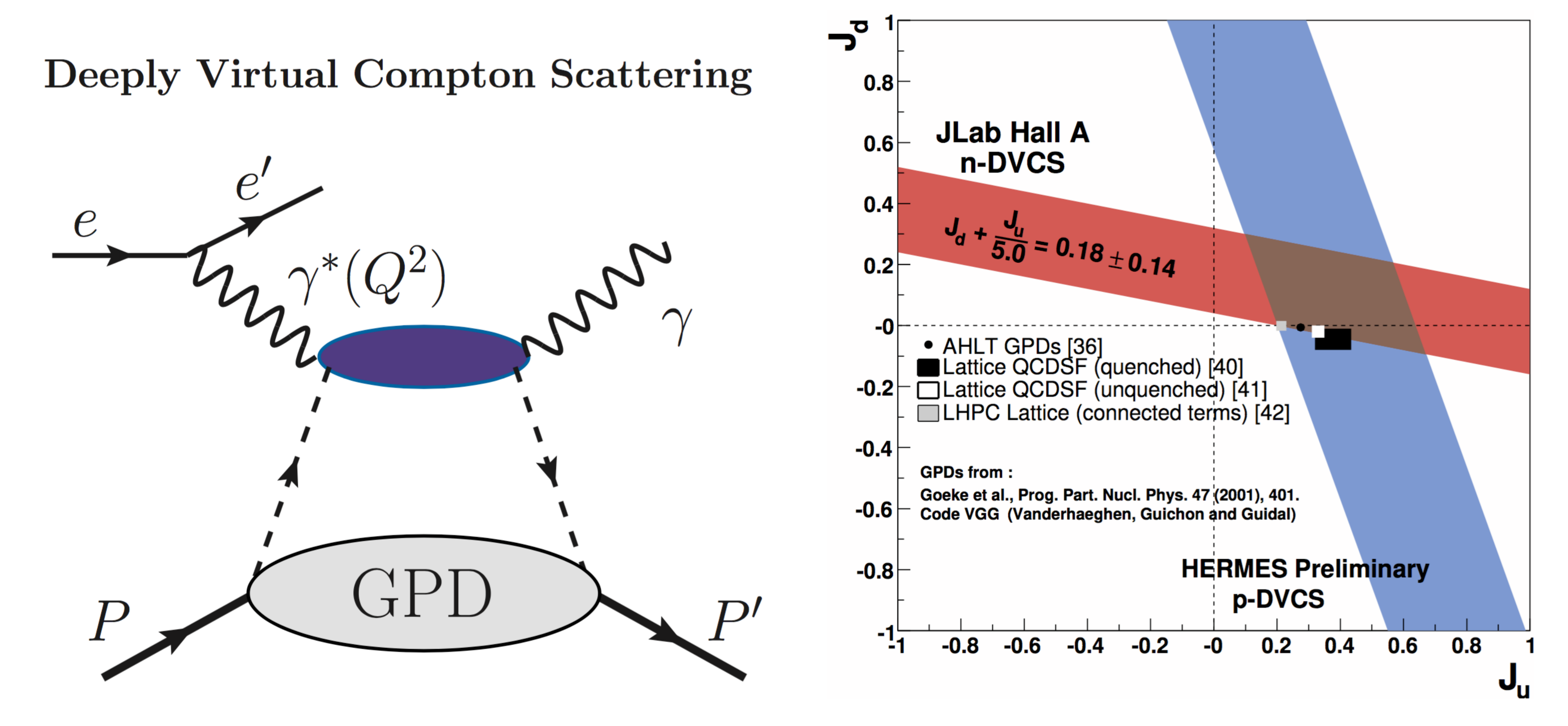}
		\caption{Extensive investigations of a new experimental process called
deeply virtual Compton scattering (DVCS)~\cite{Ji:1996nm}, shown in the left, have provided a novel approach to study the quark orbital angular momentum in proton.
In the DVCS process, an electron scatters off the nucleon with momentum $P$ and produces a high momentum real photon
and a recoiling proton with momentum $P'$. DVCS probes the generalized parton distributions which are sensitive to the orbital angular momenta of quarks and gluons.
An example (right plot) is shown from JLab Hall A analysis of spin asymmetries in DVCS and the model-dependent constraints on the up and down quark total AM. Source: Ref.~\cite{Mazouz:2007aa}.}
		\label{fig:jlabgpd}
	\end{center}
\end{figure}

It was found that the total quark (gluon) contribution to the proton spin
(also the form factor of the QCD energy-momentum tensor)
can be obtained from the moments of generalized parton
distributions (GPD's)~\cite{Ji:1996ek},
\begin{equation}
J_q=\frac{1}{2}\Delta \Sigma_q+L_q=\lim_{t,\xi\to 0}\frac{1}{2}\int dx\
x\left[H^q(x,\xi,t)+E^q(x,\xi,t)\right] \ .
\end{equation}
After subtracting the helicity contribution $\Delta\Sigma_q$ measured from
inclusive and semi-inclusive DIS experiments, the above equation will provide the
quark OAM contribution to the proton spin.
The GPD's can be measured in a new class of experiments called deep-exclusive
processes, for example, deeply virtual Compton scattering (DVCS)
and deeply virtual meson production (DVMP)~\cite{Ji:1996nm,Ji:1996ek,Collins:1996fb,Mankiewicz:1997bk}.
Both DVCS and DVMP processes belong to exclusive hard scattering processes in lepton-nucleon collisions. For example, in the DVCS process, as shown in Fig.~\ref{fig:jlabgpd}, an incoming lepton scatters off the nucleon with momentum $P$ and produces a high momentum real photon, and the recoiling nucleon with momentum $P'$. In this way,
the quark spatial position and momentum can be sampled simultaneously.
Review articles for GPDs and DVCS can be found in Ref.~\cite{Ji:1998pc,Diehl:2003ny,Ji:2004gf,Belitsky:2005qn}

Experimental efforts in these new processes
have been made at various facilities, including HERMES at DESY~\cite{Airapetian:2008aa}, Jefferson Lab 6 GeV~\cite{Mazouz:2007aa},
and COMPASS at CERN~\cite{Akhunzyanov:2018nut}.
In real photon exclusive production, the DVCS amplitude has interference with the Bethe-Heitler (BH) amplitude. This will, on the one hand, complicate the analysis of the cross section, and on the other hand, provide unique opportunities to direct access the DVCS amplitude through the interference. To obtain the constraints on the quark OAMs from these experiments, we need to find the observables which are sensitive to the GPD $E$'s. Experiments on the DVCS  from JLab Hall A~\cite{Mazouz:2007aa} and  HERMES at DESY~\cite{Airapetian:2008aa} have shown strong sensitivity to the quark OAMs in nucleon, see, e.g., Fig.~\ref{fig:jlabgpd}. In these experiments, the single spin asymmetries associated with beam or target in DVCS processes are measured, including the beam (lepton) single spin asymmetry and (target) nucleon single spin (transverse or longitudinal) asymmetries.

JLab 12 GeV facility has just started its experimental program. Multiple experiments
on DVCS and DVMP have been planned in three experimental halls. One expects
a new generation of precision data for extracting quark GPD's. From the phenomenology side,
we need to construct more sophisticated parametrizations for the GPD's.
In particular, in light of JLab experiments in next decade and future experiments
at the EIC, a rigorous and collaborative approach has to be taken
to perform the analysis of a large body of experimental data.

\subsection{Prospects of the proton spin at EIC}

\begin{figure}[htb]
		\centering
		\includegraphics[width=0.9\textwidth]{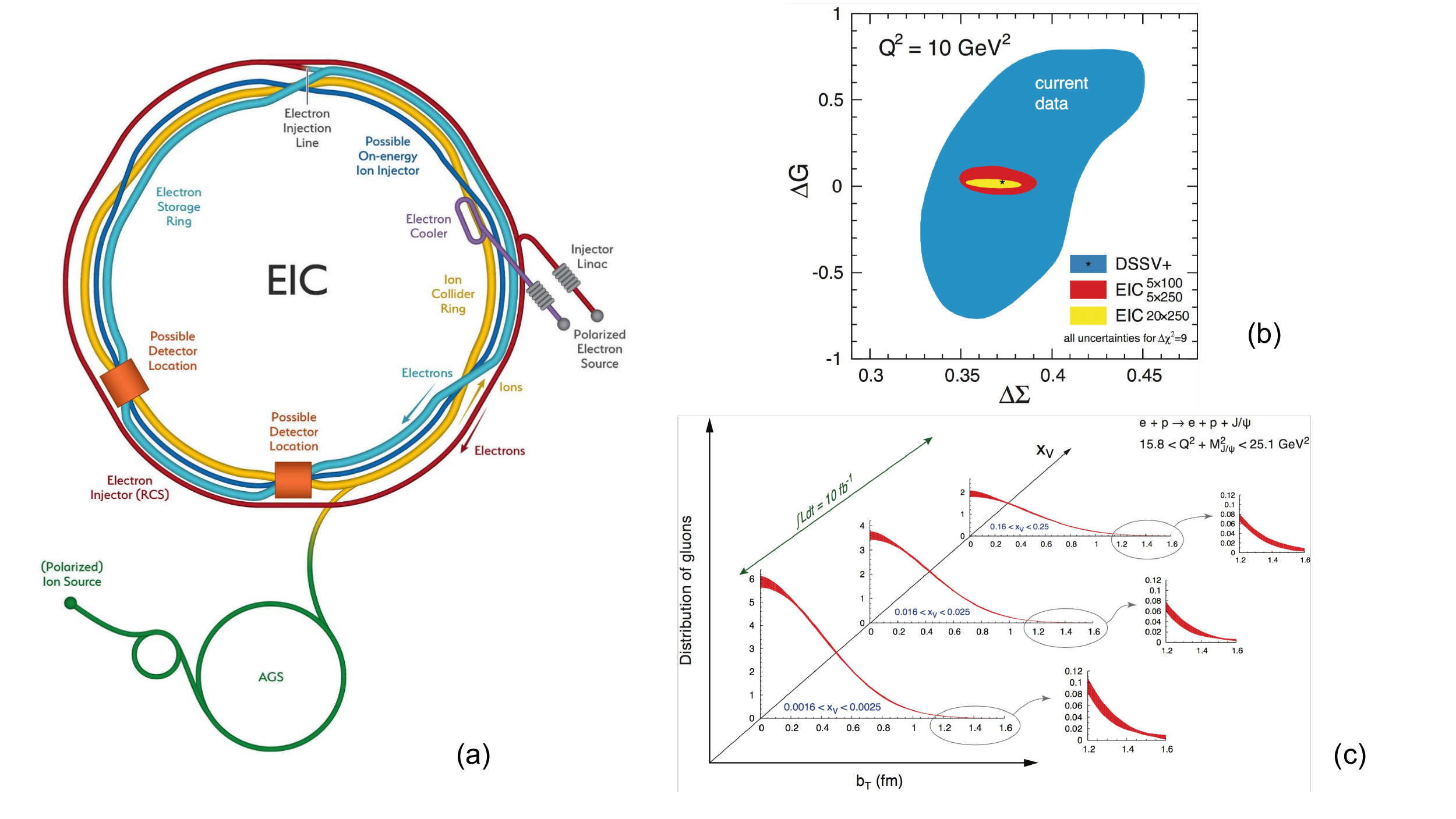}
		\caption{The planned Electron-Ion Collider (EIC) at BNL, New York, USA (Left, source: BNL). Highlights of the EIC impacts on our understanding of nucleon spin: precision on the total quark/gluon helicity contributions to the proton spin (upper right); gluon density in the transverse plane for different gluon parton $x$ from
deeply-virtual $J/\psi$ production (lower right).
This process can probe the generalized parton distribution for the gluon at small-$x$. The Fourier transform respect to the momentum transfer leads to the transverse spatial distribution of gluons in nucleon. High precision measurement of this process at the EIC will provide strong constraint on this tomography imaging. Source: Ref.~\cite{Accardi:2012qut}.}
		\label{fig:EIC}
\end{figure}

In early 2020, the DOE announced that the next major facility for nuclear physics in the US will be a high-energy and high-luminosity polarized EIC to be built at BNL.
EIC will be the first polarized electron-proton collider and the first electron-nucleus collider as well.
The primary goal of the EIC is to precisely image gluon distributions in nucleons and nuclei, to reveal the origins of the nucleon mass and spin, and to explore the new QCD frontier of cold nuclear matter~\cite{Accardi:2012qut,Boer:2011fh}. EIC will impact our understanding of nucleon spin in many different ways. In the following, we highlight some of the most obvious ones:

\begin{itemize}

\item  The quark and gluon helicity contributions to the proton spin are among the major emphases of the planned EIC. With the unique coverage in both $x$ and $Q^2$, it will provide the most stringent constraints on $\Delta \Sigma$ and $\Delta G$~\cite{Accardi:2012qut}. Shown in Fig.~\ref{fig:EIC} is the possible reduction in their uncertainties with the proposed EIC. Clearly, it will make a huge impact on our knowledge of these quantities, unparalleled by any other existing or anticipated facility.

\item There will be a comprehensive research program on gluon GPD's at the EIC. Apart from
 providing the first hand constraints on the total quark/gluon AM contributions to the proton spin,
  the GPD's provide important information on the nucleon tomography, for example, the 3D imaging of partons inside the proton~\cite{Burkardt:2002hr,Belitsky:2003nz}. With wide kinematic coverage at the EIC, a particular example shown in Fig.~\ref{fig:EIC} is that the transverse imaging of the gluon can be precisely mapped out from the detailed measurement of hard exclusive $J/\psi$ production. Together with the gravitational form factors extracted from the DVCS, this will provide unprecedented exploration of nucleon tomography and deepen our understanding of the nucleon spin structure in return. Pioneer experimental effort to constrain the gravitation form factor from DVCS experiment at JLab has been carried out in Ref.~\cite{Burkert:2018bqq}.
	
\item The EIC may shed light on the quark/gluon canonical OAM directly through various hard diffractive processes. A particular example has been studied recently in Refs.~\cite{Ji:2016jgn,Hatta:2016aoc}. Here, by applying the connection between the parton Wigner distribution and OAM~\cite{Hatta:2011ku,Ji:2012sj,Lorce:2011kd,Lorce:2011ni}, one can show that the single longitudinal target-spin asymmetry in the hard diffractive dijet production is sensitive to the
    canonical gluon OAM distribution. The associated spin asymmetry leads to a characteristic azimuthal angular correlation between the proton momentum transfer and the relative transverse momentum between the quark-antiquark pair. With a hermetic detector designed for the EIC, this observable can be well studied in the future, and will help us obtain the final piece in the IMF helicity sum rule.
\end{itemize}

 An important theoretical question concerns the asymptotic small-$x$ behavior of the polarized PDFs and their contributions to the spin sum rule. There have been some progress to understand the proton spin structure at small-$x$ from the associated small-$x$ evolution equations~\cite{Kovchegov:2015pbl,Kovchegov:2016weo,Kovchegov:2018znm,Boussarie:2019icw,Tarasov:2020cwl}. More theoretical efforts are needed to resolve the controversial issues raised in these derivations. The final answer to these questions will provide important guideline for the future EIC, where proton spin structure is one of the major focuses.

\section{Conclusion}

After 30 years since the EMC publication of the polarized DIS data,
there has been much progress in understanding the spin structure.
There are two well-established approaches to look at the composition of the proton spin:
frame-independent approach (``Ji sum rule'') and infinite-momentum-frame parton approach
(``Jaffe-Manohar sum rule''). In the frame-independent approach, the quark orbital and gluon contributions
can be obtained from moments of generalized parton distributions. Results
from Jlab 6 GeV experiments and HERMES data suggest a substantial quark orbital
contribution. In the partonic picture of Jaffe and Manohar, the quark and gluon helicity have simple
physics appeal, and the result from RHIC spin has provided important
constraint on the total gluon helicity $\Delta G$. Development of LaMET along with lattice QCD
simulations provides the first-principles calculations of the spin
structure, and the first results have provided an interesting overall picture.
Jlab 12 GeV program will provide much improved data on the quark GPDs and OAM.
EIC can provide high-precision measurements on the gluon helicity $\Delta G$ and total
angular momentum contributions.

{\it Acknowledgment.}---This article is dedicated to late V. Hughes whose drive
in polarized DIS experiments lead to the surge and unabated interest in the proton spin structure,
and perhaps to the EIC project in the US. We thank C. Aidala, C. Alexandrou, M. Burkardt, Y. Hatta, D. Hertzog, R. Jaffe,
and K. F. Liu for useful communications relating to this article. This material is supported by the U.S. Department of Energy, Office of Science, Office of Nuclear Physics, under contract number DE-AC02-05CH1123, DE-SC0012704 and DE-SC0020682, and within the framework of the TMD Topical Collaboration.

\bibliography{spin}

\end{document}